\documentclass[%
a4paper,
reprint,
superscriptaddress,
showpacs,
amsmath,amssymb,
acs,
jctc,
showkeys,
onecolumn,
longbibliography
]{revtex4-2}

\setcitestyle{super}

\usepackage[skip=1pt plus 0.2pt]{parskip}
\usepackage{xcolor}
\usepackage{hyperref}
\usepackage{multirow}
\usepackage{acronym}
\usepackage{csquotes}
\usepackage{booktabs}
\usepackage{geometry}
\usepackage{graphicx}
\usepackage{braket}

\DeclareMathOperator{\e}{e} 		
\providecommand{\abs}[1]{\lvert#1\rvert}

\newcommand{\bracket}[2]{{\left\langle \vphantom{#1 #2} #1 \,\right|
		\left.\hspace{-0.15em} \vphantom{#1 #2} #2\right\rangle}}
\newcommand{\mbf}[1]{\mathbf{#1}}
\newcommand{\ra}{\ensuremath{\rightarrow}}

\newcommand*{\citen}{}
\DeclareRobustCommand*{\citen}[1]{%
	\begingroup
	\romannumeral-`\x 
	\setcitestyle{numbers}%
	\cite{#1}%
	\endgroup
}

\usepackage[normalem]{ulem}

\begin{document}

	\title{Towards real chemical accuracy on current quantum hardware through the transcorrelated method}

	\date{\today}

	\begin{abstract}

Quantum computing is emerging as a new computational paradigm with the potential to transform several research fields, including quantum chemistry. 
However, current hardware limitations (including limited coherence times, gate infidelities, and connectivity) hamper the implementation of most quantum algorithms and call for more noise-resilient solutions. 
We propose an explicitly correlated Ansatz based on the transcorrelated approach to target these major roadblocks directly. 
This method transfers -- without any approximation -- correlations from the wavefunction directly into the Hamiltonian, thus reducing the resources needed to achieve accurate results with noisy quantum devices. 
We show that the transcorrelated approach allows for shallower circuits and improves the convergence towards the complete basis set limit, providing energies within chemical accuracy to experiment with smaller basis sets and, thus, fewer qubits. 
We demonstrate our method by computing bond lengths, dissociation energies, and vibrational frequencies close to experimental results for the hydrogen dimer and lithium hydride using two and four qubits, respectively. 
To demonstrate our approach's current and near-term potential, we perform hardware experiments, where our results confirm that the transcorrelated method paves the way toward accurate quantum chemistry calculations already on today's quantum hardware.

	\end{abstract}
	
	\author{Werner Dobrautz}
	\email{dobrautz@chalmers.se}
\thanks{These two authors contributed equally}
	\affiliation{
		Department of Chemistry and Chemical Engineering, 
		Chalmers University of Technology, 41296 Gothenburg, Sweden
	}
	
	\author{Igor O. Sokolov}
	\thanks{Current address: PASQAL, 7 rue Léonard de Vinci, 91300 Massy, France}
\thanks{These two authors contributed equally}

	\affiliation{IBM Quantum, IBM Research Zurich, Säumerstrasse 4, 8803 Rüschlikon, Switzerland}
	
	\author{Ke Liao}
	\thanks{Current address: Division of Chemistry and Chemical Engineering, California Institute of Technology, Pasadena, USA 91125}
	\affiliation{Max Planck Institute for Solid State Research, Heisenbergstr. 1, 70569 Stuttgart, Germany}
	
	\author{Pablo L\'opez R\'ios}
	\affiliation{Max Planck Institute for Solid State Research, Heisenbergstr. 1, 70569 Stuttgart, Germany}%
	
	\author{Martin Rahm}
	\affiliation{
		Department of Chemistry and Chemical Engineering, 
		Chalmers University of Technology, 41296 Gothenburg, Sweden
	}
	
	\author{Ali Alavi}
	\affiliation{Max Planck Institute for Solid State Research, Heisenbergstr. 1, 70569 Stuttgart, Germany}%
	\affiliation{Yusuf Hamied Department of Chemistry, University of Cambridge, Lensfield Road, Cambridge CB2 1EW, United Kingdom}%
	
	\author{Ivano Tavernelli}
	\email{ita@zurich.ibm.com}
	\affiliation{IBM Quantum, IBM Research Zurich, Säumerstrasse 4, 8803 Rüschlikon, Switzerland}

	\maketitle
	
	\section{\label{sec:intro} Introduction}
	
	Quantum computing~\cite{Feynman1982, Benioff_1980} has the potential of providing a significant speedup in the simulation of natural sciences 
	compared to classical computational approaches.
	However, the implementation and application of quantum algorithms to relevant problems, e.g., in electronic structure theory, is still in its infancy. 
	In this work, we show that the solution of molecular electronic structure problems using an explicitly correlated approach based on the transcorrelated (TC) method~\cite{Hirschfelder1963, Boys1969, BoysHandy1969, Handy1969, Handy1969b} can take advantage of a quantum computing implementation. 
	In fact, by enabling accurate and affordable quantum chemistry calculations for relevant problems, we argue that TC can become the method of choice for near-term demonstration of quantum advantage with state-of-the-art, noisy quantum computers. 
	
	Computational quantum chemistry (QC) is concerned with the solution of the electronic Schr\"odinger equation (SE) to obtain ground and excited state wavefunctions, their energies, and corresponding molecular properties.~\cite{Helgaker_2000}
	Sufficiently accurate modeling of the correlated motion of electrons would allow the description of many groundbreaking yet unsolved physical and chemical phenomena, such as unconventional high-$T_c$ superconductivity,~\cite{Bednorz_1986} photosynthesis,~\cite{Renger1987} and nitrogen fixation.~\cite{Holm1996}
	More generally, an efficient solver for the SE will make it possible to predict and design materials with novel and improved chemical and physical properties.
	
\begin{figure*}
		\centering
		\includegraphics[width=0.97\textwidth]{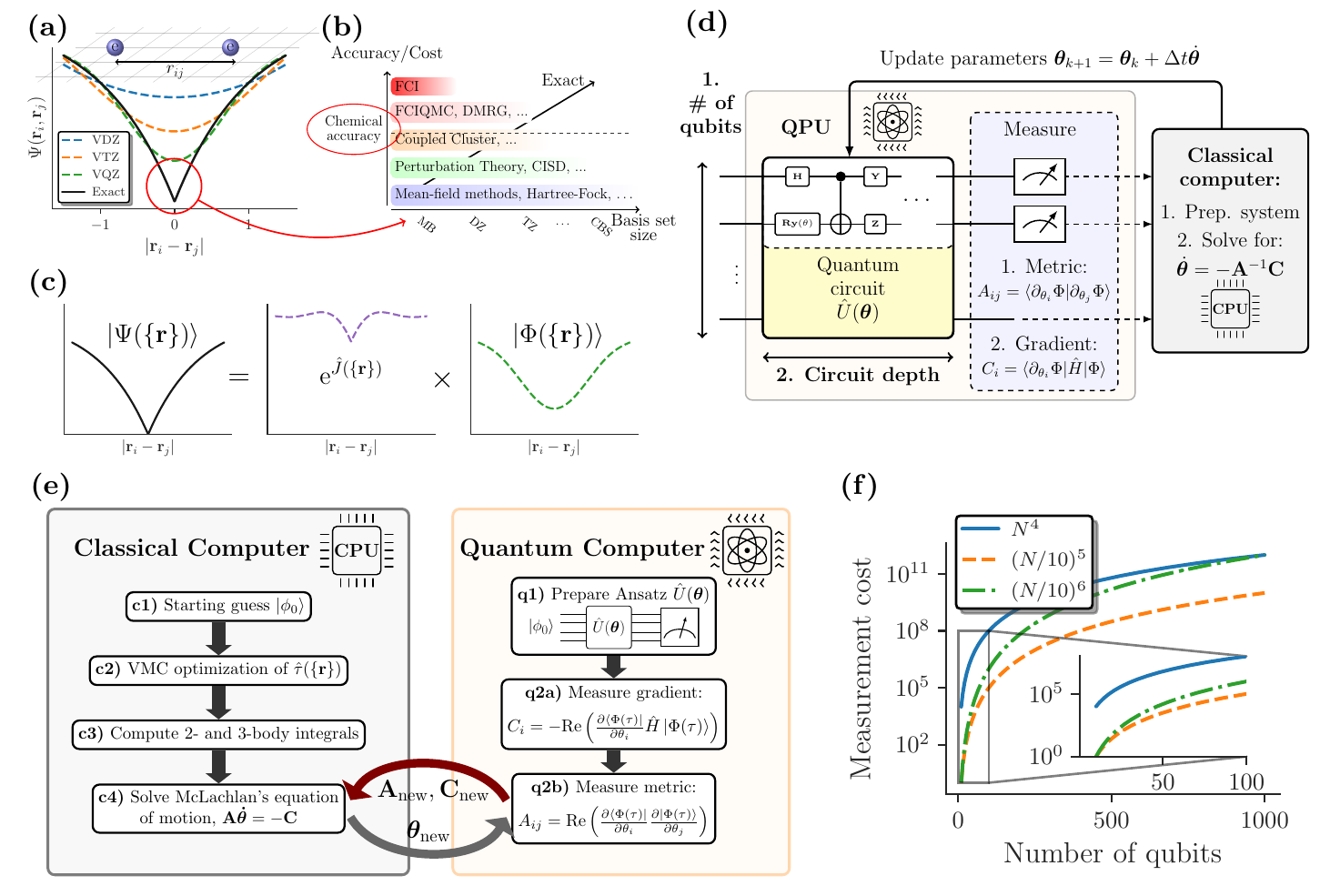}
		\vspace*{-0.4cm}
		\caption{\label{fig:sketch-comb}
			\textbf{(a)} Electronic cusp and increasing basis size to capture sharp features of the exact wavefunction.
			\textbf{(b)} Hierarchy of methods and basis size towards exact CBS solution.
			\textbf{(c)} Jastrow Ansatz to capture the cusp feature, leaving a smoother $\ket{\Phi(\{\mbf r\})}$ to solve for.
			\textbf{(d)} VarQITE algorithm, where the metric, $\mathbf{A}$, and the gradient, $\mathbf{C}$, are measured on the QPU.  The linear system of equations $\mathbf{A}\dot{\boldsymbol{\theta}} = -\mathbf{C}$ is solved on a CPU to obtain the new parameters, which are fed back to the QPU.
			\textbf{(e)} Workflow of the TC-VarQITE approach to solve for the right eigenvector and groundstate energy of the TC Hamiltonian. On a CPU, we: (c1) Perform a Hartree-Fock in a chosen basis set  and optionally a MP2 calculation (using PySCF\cite{Sun2017, Sun2020} or OpenMolcas\cite{LiManni2023} in this work) to obtain starting orbitals and $\ket{\phi_0}$.
			(c2) Optimize the Jastrow factor with VMC.
			(c3) Compute the 2- and 3-body integrals 
			for the subsequent electronic structure calculation. 
			Then we enter the quantum-classical optimization loop, sketched in \textbf{(d)}, consisting of: (q1) Preparing a parametrized Ansatz and measuring the gradient (q2a) and the metric (q2b).
			\textbf{(f)} 
			Measurements of operators containing 2-body and 3-body terms, scale as $N^4$ (solid blue line) and $N^6$,  respectively, where $N$ is the number of orbitals in the basis set. The TC method reduces the number of necessary orbitals by about an order of magnitude (green dash-dotted line). 
			Neglecting 3-body excitations with six unique indices~\protect\cite{Giner2021, Dobrautz2022} or mean-field approximations~\cite{Christlmaier2023} reduces the scaling of the TC method to the fifth or even the fourth power of $N$ (with negligible errors in the systems studied in this work).
		}
	\end{figure*}

	A wide variety of approximate QC computational approaches have been developed, ranging from inexpensive mean-field Hartree-Fock (HF)~\cite{Helgaker_2000} to more reliable but expensive density matrix renormalization group (DMRG),~\cite{White1992} coupled cluster (CC)~\cite{Cicek1966} and quantum Monte Carlo (QMC) methods.~\cite{Nightingale1998} At their limit, i.e., in the absence of truncation and approximations, several such methods can approach the exact result, named the full configuration interaction (FCI) solution. 
	FCI scales combinatorially with the number of electrons in a syste
	and the size of the utilized \emph{basis set expansion}, see Fig.~\ref{fig:sketch-comb}a-b.
	
	The accuracy of typical quantum chemistry calculations is strongly affected by the quality of the \emph{basis set}, which is used to expand the many-electron wavefunction in terms of one-electron basis functions.~\cite{Helgaker_2000} 
	Such functions are commonly smooth Gaussian-type orbitals (GTOs),~\cite{Ditchfield_1971, Dunning_1989} that produce tractable one- and two-body integrals, but fail to capture the electron-cusp condition.~\cite{Kato1957}
	The cusp condition is a sharp feature of the exact ground state wavefunction induced by the divergence of the Coulomb potential at electron coalescence, which can typically only be captured through large basis sets. 
	Using a larger number of basis functions results in a sizeable increase in the required computational resources (see Fig.~\ref{fig:sketch-comb}a and b). 
	Thus, more accurate methods are practically limited to small problem sizes even when using high-performance computing resources.
	
	Quantum processors, on the other hand, harness quantum mechanical phenomena to potentially enable a significant leap in computation.~\cite{Nielsen2012}
	By using \emph{quantum bits} (qubits) as the basic unit of information and computation, quantum computers can encode  exponentially growing problem sizes, $2^n$, into the Hilbert space of $n$ qubits.
	Specifically designed quantum algorithms can then leverage wave function superposition and entanglement to solve certain classically challenging problems.\cite{mcardle2020quantum}
	Despite this potential, the sizes of quantum chemistry systems treatable on current, noisy quantum hardware are still relatively modest and do not yet exceed the capability of conventional computing approaches.
	The main challenges to solve are qubit decoherence, gate noise, and the limited number of available qubits as, in the case of quantum chemistry, the number of qubits 
	scales with the size of the required basis set. 
	Thus, many methods to reduce the number of necessary qubits have been recently proposed. 
	Among others, there are approaches leveraging system symmetries, others based on concepts such as entanglement forging,~\cite{Eddins2022} tensor hypercontraction,~\cite{Lee2021} low-rank representations,~\cite{Motta_2021}  methods for reducing the basis set size~\cite{Motta2020}
	using Daubechies wavelets,~\cite{Hong2022}
	explore basis-set-free solutions,~\cite{Kottmann2021}
	or Hamiltonian downfolding techniques.~\cite{Huang2023, Bauman2019, Bauman2019b, NicholasP2021, Bauman2022}

	Explicitly correlated methods,~\cite{Hylleraas1929, Httig2011, Kong2011, Tenno2011, Tenno2012, Grneis2017} like 
	the R12\cite{Kutzelnigg1985} or F12 approaches,\cite{Tenno2004, Tenno2004b}
	can reduce the need for large basis set expansions by directly incorporating the electronic cusp condition
	in the wavefunction Ansatz. 
	Recently, it has been shown that methods based on these explicitly correlated approaches can yield accurate results already with relatively small basis sets and thus reduce the number of necessary qubits on quantum hardware.~\cite{Motta2020, schleich2021improving, Mcardle2020, Sokolov2022, Kumar2022}
		Motta \textit{et al.}\cite{Motta2020} have used canonical transcorrelated F12 (CT-F12) theory\cite{Yanai2006, Neuscamman2010, Yanai2012} to study several small molecular species, requiring far less quantum resource than conventional approaches. 	
		Kumar \textit{et al.}\cite{Kumar2022} extended CT-F12 to obtain accurate excited state energies with reduced quantum resources, and Schleich \textit{et al.} used [2]$_{\rm R12}$ theory\cite{Torheyden2009} to a posteriori correct energy estimates obtained on quantum hardware. 
	
	In the TC approach~\cite{Boys1969, Handy1969, Handy1969b, Dobrautz2019, Cohen2019, Guther2021, Baiardi2020, Baiardi2022, Liao2023, Liao2021, Schraivogel2023, Schraivogel2021, Ammar2022, Tenno2023} a correlated Ansatz -- exactly incorporating the cusp condition -- is applied and used to perform a similarity transformation of the electronic Hamiltonian, $\hat H$, describing the \emph{ab initio} chemical system.
	The undisputed benefit of the TC method is that it yields highly accurate results with very small basis set expansions~\cite{Cohen2019, Dobrautz2022, Giner2021} and thus reduces the number of required qubits as well as the circuit depth on a quantum computer. 
	The reduced circuit depth arises because the TC Hamiltonian has a more compact ground state,~\cite{Dobrautz2019, Mcardle2020, Sokolov2022} which can be accurately represented with shallower circuits. 
	
	The main challenge concerning implementing the TC approach is that the corresponding Hamiltonian is non-Hermitian.
	Most near-term quantum computing approaches rely on the minimization of the expectation value of a Hermitian operator (i.e., the energy as the expectation value of the Hamiltonian $\hat H$) using the variational quantum eigensolver (VQE)~\cite{peruzzo_variational_2014}.
	To overcome this limitation, in this work, we use a variational Ansatz-based formulation of the projective quantum imaginary-time evolution (QITE),~\cite{Motta_2019, Gomes2020, Cao2022, Nishi2021, Tsuchimochi2023} namely the variational QITE (VarQITE) algorithm.~\cite{Yuan_2019, mcardle2019variational}
	With minor modifications, VarQITE enables the study of non-Hermitian problems,~\cite{Mcardle2020, Sokolov2022} such as optimizing the TC Ansatz in a quantum computing setting. 
	
		The main differences between CT-F12 and the TC approach are: 
		\textbf{(a)} CT-F12 uses a unitary operator in the similarity
		transformation, which does not terminate naturally. Consequently, however, the transformed Hamiltonian remains Hermitian.   
		\textbf{(b)} Two major approximations are used in CT-F12 theory. 
		The Baker-Campbell-Hausdorff (BCH) expansion of the similarity transformation is truncated at the second order, and in the double commutator term, an effective 1-body Fock operator is used instead of the full Hamiltonian. 
		The benefits of CT-F12 is that the Hamiltonian remains Hermitian and contains only up to 2-body terms. Additionally, by using a F12 Slater-type geminal,\cite{Tenno2004} the spin dependence of the electron-electron cusp\cite{Pack1966} can be taken into account.\cite{Zhang2012}
		Drawbacks include \textbf{i)} that the BCH expansion in CT-F12 is truncated at the second commutator and any higher-than-two-body interactions are ignored.
		This truncation induces errors that are not easy to eliminate, especially in case of strong static correlations;~\cite{Neuscamman2010,kowalski2023b} \textbf{ii)} the need to use a projection to ensure the orthogonality between the small and the augmented basis set, which can make the use of more sophisticated correlators, as used in our study, very complicated. 
		As a result, CT-F12 uses a simpler correlated Ansatz and does not correct 1-body particle incompleteness,\cite{Kumar2022}, which can lead to worse results compared to TC approaches. The 1-body particle incompleteness was recently addressed in the work by Kumar \textit{et al.}~\cite{Kumar2022} 
		Opposed to the TC and CT-F12 approach, [2]$_{\rm R12}$ is an \textit{a posteriori} correction, where one- and two-particle reduced density matrices, measured on quantum hardware, are used to improve final energy estimates.

	To date, the largest quantum chemistry calculations performed on real quantum computers include Hartree-Fock calculations of a 12-atom hydrogen chain and a diazene isomerization~\cite{Arute2020} along with correlated calculations of BeH$_2$~\cite{Kandala2017} and H$_2$O.~\cite{Nam_2020}
	The primary purpose of these calculations was to showcase the proof-of-concept of quantum computing using small basis sets rather than accuracy. 
	On the contrary, the TC method will pave the way  toward accurate quantum chemistry calculations on quantum computers, allowing for precise calculations close to the complete basis set (CBS) limit using small basis sets. 
		Although explicitly correlated approaches, such as TC, significantly reduce the resource requirements for conventional methods like FCIQMC or DMRG, practical problem sizes still quickly surpass those that these approaches can handle.
	We, therefore, believe that by reducing the number of qubits and gate operations, the TC approach will make reliable quantum chemistry calculations of relevant problems with current and future error-mitigated quantum devices possible.
	Potentially surpassing the capabilities of conventional approaches in the future. 
	
	\section{\label{sec:theory} Theory and Algorithms}
	
	Within the Born-Oppenheimer approximation, the molecular Hamiltonian in first quantization (in atomic units, $\hbar = \abs{e} = m_e = 4\pi\epsilon_0 = 1$) is given by
	\begin{equation}\label{eq:hamil}
		\hat H = -\sum_i^{n_e} \left( \frac{1}{2} \nabla_i^2 + \sum_a^{N_a}  \frac{Z_a}{\abs{\mbf r_i - \mbf R_a}}\right) + \sum_{i<j} \frac{1}{\abs{\mbf r_i - \mbf r_j}} \, .
	\end{equation}
	In Eq.\eqref{eq:hamil}, $n_e$ is the number of electrons, $N_a$ the number of nuclei, $\mbf R_a$ and $Z_a$ the position and atomic number of nucleus $a$, and  $\mbf r_i$ the position of electron $i$.
	The divergence of the Coulomb potential, $\frac{1}{r_{ij}}$, induces a sharp cusp-like feature of the exact electronic wavefunction, $\ket{\Psi_0(\mbf r)}$, at electron coalescence, $r_{ij} = \abs{\mbf r_i - \mbf r_j} = 0$~\cite{Kato1957} (Fig.~\ref{fig:sketch-comb}a).
	This sharp feature of $\ket{\Psi_0(\mbf r)}$ is challenging to capture using basis functions based on smooth GTOs and requires the use of large basis sets for accurate quantum chemical results (Fig.~\ref{fig:sketch-comb}b).
	
	By introducing an explicit dependence on the electron-electron distances into the wavefunction via a Jastrow Ansatz,~\cite{Jastrow1955} $\ket{\Psi} = \e^{\hat J} \ket{\Phi}$, it is possible to exactly describe the non-smooth behavior of $\ket{\Psi}$, while leaving a much smoother wavefunction, $\ket{\Phi}$, to solve for (Fig.~\ref{fig:sketch-comb}c).
	$\hat J = \hat J (\mbf r_1, \dots, \mbf r_n)$ is an optimizable correlator depending on the positions of the electrons.
	The TC method~\cite{Boys1969} incorporates this correlated Ansatz directly into the Hamiltonian of the system via a similarity transformation,
	
	\begin{align}\label{eq:tc}
		&\stackrel{\text{Original:}}{\hat H \ket\Psi = E\ket{\Psi}}, \quad \text{with}\; 
		\stackrel{\text{Correlated Ansatz:}}{\ket{\Psi} = \e^{\hat J} \ket{\Phi}}\quad \\[5pt]
		&	 \ra 
		\quad 
		\stackrel{\text{Transcorrelated Problem:}}{
			\bar{H}
			\ket{\Phi} = E \ket{\Phi},\; \bar{H} =  \e^{-\hat{J}} \hat H \e^{\hat{J}}}. \label{eq:tc-2}
	\end{align}
	This similarity transformation removes the Coulomb singularity of the original molecular Hamiltonian, Eq.\eqref{eq:hamil},~\cite{Luo2018} and consequently increases the smoothness of the sought-after ground state wavefunction $\ket{\Phi}$.\cite{Fournais2005}
	For the molecular Hamiltonian, Eq.~\eqref{eq:hamil}, the TC Hamiltonian, $\bar{H} =  \e^{-\hat{J}} \hat H \e^{\hat{J}}$, can be calculated exactly.~\cite{Cohen2019}
	$\bar{H}$ possesses non-Hermitian two-body and additional three-body interaction terms.

	In our applications, we use a Drummond-Towler-Needs Jastrow factor,~\cite{Drummond2004, Lopez2012} which we optimize with variational Monte Carlo (VMC)\cite{Ceperley1977, Foulkes2001, Haupt2023} (with an efficient scaling of $\mathcal{O}(n_e^3)$ on conventional hardware) using the \texttt{CASINO} package.~\cite{Needs2020, LopezRios2006}
	We optimize the Jastrow factor by minimizing the variance of the TC reference energy, as proposed recently in Ref.~\cite{Haupt2023}
	We then use the transcorrelated Hamiltonian integral (\texttt{TCHint}) library to calculate the 2- and 3-body integrals required to construct the molecular Hamiltonian in second quantization (Fig.~\ref{fig:sketch-comb}e).
	We want to point the interested reader to the recent Ref.~\cite{Haupt2023}, the Methods Section\ref{sec:methods} and the Supporting Information (SI)~\cite{SI} for more details. 
	Sample input files of the VMC optimization and integral calculation can be found in the \texttt{Github} repository accompanying the paper.\cite{github-repo}
	
	Due to being non-Hermitian, the TC Hamiltonian, depending on the VMC optimization of the parameters in the Jastrow Ansatz, can theoretically yield energies below the exact CBS limit result in a finite basis.
	However, since the TC Hamiltonian originates from a similarity transformation, 
	the correct eigenvalues are obtained when using a large enough basis (approaching the CBS limit). 
	The issue of non-variationality  has been thoroughly studied in the recent Ref.~\cite{Haupt2023} for the Jastrow factors and VMC optimization used in this work. 
	It has been found that the when optimizing the Jastrow factor by minimizing the variance of the TC reference energy, as done in this work, 
	the results usually converge
	to the CBS limit 
	from above.
	Additionally, in this and all recent studies using the TC approach (combined with a variety of methods and applied to different types of problems),~\cite{Dobrautz2019, Cohen2019, Ammar2022, Haupt2023, Schraivogel2021, Schraivogel2023, Christlmaier2023}
	the amount by which the TC results falls below the CBS limit is 
	in all cases small enough to be safely ignored in practice.

	Due to being non-Hermitian, $\bar{H}$ has different left ($\bra{\Phi^L_i} E_i = \bra{\Phi^L_i}\bar{H}$) and right ($\bar H \ket{\Phi^R_i} = E_i\ket{\Phi_i^R}$), eigenvector solutions, which form a bi-orthonormal basis, $\braket{\Phi_i^L\vert \Phi_j^R} = \delta_{i,j}$.
		Another consequence of the non-Hermiticity of the TC Hamiltonian is that 
		VQE can not be used to solve for the ground state 
		as the variational principle does not apply.
		To circumvent this obstacle, 
	in this work, we solve for the right ground state wavefunction, $\ket{\Phi_0^R}$, (we drop the superscript \enquote{$R$} from here on), with the 
	VarQITE algorithm.~\cite{mcardle2019variational, Mcardle2020}
	An additional benefit of the TC method is that the right ground-eigenvector of $\bar H$, $\ket{\Phi_0}$, has a more compact form compared to the non-TC ground state solution, $\ket{\Psi_0}$.~\cite{Dobrautz2019, Sokolov2022}
	Consequently, $\ket{\Phi_0}$ is easier to prepare on quantum hardware with shallower circuits.~\cite{Sokolov2022}

	QITE can be recast into a hybrid quantum-classical variational form (VarQITE)~\cite{mcardle2019variational, Mcardle2020} (Fig.~\ref{fig:sketch-comb}d), obtained by applying McLachlan's variational principle to the imaginary-time SE,
	\begin{equation}\label{eq:mclachlan}
		\delta \abs{\abs{\left(\partial/\partial \tau + \bar H - E_{\tau}\right)\ket{\Phi(\tau)}}} = 0,
	\end{equation}
	where $\tau = it$ is imaginary time, $\abs{\abs{\ket{\Phi}}} = \sqrt{\bracket{\Phi}{\Phi}}$ is the norm of a quantum state $\ket{\Phi}$ and $E_{\tau} = \bra{\Phi(\tau)}\bar H\ket{\Phi(\tau)}$ is the expected energy at time $\tau$.
	With a parametrized circuit Ansatz, $\hat U (\boldsymbol{\theta}(\tau))$, with $n_\theta$ parameters, to represent/approximate the target right eigenvector, $\hat U (\boldsymbol{\theta}(\tau)) |\phi_0\rangle=\ket{\Phi(\tau)}$, of the TC Hamiltonian, Eq.\eqref{eq:mclachlan} leads to a linear system of equations
	\begin{equation}\label{eq:lse}
		\mbf A \dot{\boldsymbol{\theta}} = - \mbf C \, ,
	\end{equation}
	which is solved on a classical computer.
	The updated parameters are obtained from $\boldsymbol{\theta}(\tau + \Delta\tau) = \boldsymbol{\theta}(\tau) + \Delta\tau \dot{\boldsymbol{\theta}}$ for a chosen time-step, $\Delta\tau$.
	The vector $\mbf C$ is composed of energy gradients while the matrix $\mbf A$ is related to the quantum Fischer information matrix or Fubini-Study metric~\cite{Stokes2020} and encodes the metric in parameter space of the Ansatz $\hat U(\boldsymbol{\theta})$. 
	The VarQITE method can circumvent potential parameter optimization pitfalls,~\cite{Stokes2020, McClean2018} by a deterministic update of the circuit parameters according to Eq.\eqref{eq:lse}. 
	Both quantities $\mbf C$ and $\mbf A$ in Eq.~\eqref{eq:lse} are sampled from the quantum circuit. 
	This comes at the cost of $\mathcal{O}(n_{\theta}^2)$ circuit evaluations to measure the matrix $\mbf A$ at each iteration. 
	However, accurate approximations~\cite{Gacon2023} have been proposed, which reduce the measurement scaling to linear,~\cite{Stokes2020, Fitzek2023} or even constant scaling,~\cite{Gacon2021} and it was recently shown by van Straaten and Koczor~\cite{Straaten2021} that the measurement cost of the gradient will dominate for large-scale quantum chemistry applications.
	The implementation of the VarQITE algorithm and the necessary modifications for non-Hermitian Hamiltonians (TC-VarQITE) are detailed in the Methods Section~\ref{sec:methods}.

	The evaluation of the 3-body terms of the TC Hamiltonian might raise the question of scalability, as a 3-body term requires $\mathcal{O}(N^6)$ measurements, where $N$ is the number of basis functions/(spin-)orbitals in the basis set.
	However, one should consider that for an efficient implementation of the TC-VarQITE algorithm, one does not need an accurate evaluation of the energy (with all 3-body terms) at each iteration until convergence is reached. 
	This can be monitored by calculating the norm $||\mbf A^{-1} \mbf C||$.
	Furthermore, as the TC approaches enables a faster convergence towards the basis set limit than conventional approaches, we expect an overall decrease in the number of orbitals $N$ (and thus qubits) by an order of magnitude, leading to a $\mathcal{O}[(N/10)^6]$ scaling, i.e., to a decrease of the pre-factor by six orders of magnitude. 
	Overall, this implies that in the regime up to 1000 qubits, the TC-VarQITE method entails orders of magnitude fewer measurements than in the non-TC case (see Fig.~\ref{fig:sketch-comb}f).
	Furthermore, recent studies~\cite{Dobrautz2022, Haupt2023, Christlmaier2023} show that the number of terms in the TC Hamiltonian can be reduced to $(N/10)^5$ (by neglecting 3-body excitations with six unique indices\cite{Dobrautz2022}) or even to $(N/10)^4$ by
	neglecting the pure normal ordered\cite{Kutzelnigg1997} three-body operators and incorporating the remaining 3-body contributions in the two-, one-, and zero-body integrals~\cite{Christlmaier2023} 
	(shifting the crossover far beyond 1,000 qubits). 
	The applicability of these types of approximations has to be carefully considered for each studied system. However, 
	the $N^4$-scaling method introduced in Ref.~\cite{Christlmaier2023}
	has  recently been applied to the entire \enquote*{HEAT} benchmark set~\cite{Tajti2004} and the $N^5$ scaling approximation has been used in Ref.~\cite{Dobrautz2022} for all first-row atoms, as well as the molecular systems CH$_2$, FH and H$_2$O, and in Ref.~\cite{Schraivogel2023} for relative energies of molecular systems.
	
		According to work by Loaiza \textit{et al.},\cite{Loaiza2023}
		the 1-norm, $\sum_i \vert c_i \vert$ of the coefficients, $c_i$, of the linear combination of unitaries decompositions for molecular electronic structure Hamiltonians,
		\begin{equation}\label{eq:lcu}
			\hat H = \sum c_i \hat P_i,
		\end{equation}
		is the main figure of merit associated with the quantum circuit complexity to measure the Hamiltonian. 
		We measured the 1-norm of the coefficients corresponding to diagonal, one-, two- and three-body operators of the LiH (TC) Hamiltonian and compiled the results in Table~S9 in the SI\cite{SI}.
		We found that the normalized 1-norm of the 3-body operators, $\sum_{i,
			\textrm{3-body}} \vert c_i\vert / \sum_i \vert c_i\vert$, is substantially smaller (below 0.1 \%) than the remaining contributions in the qubit Hamiltonians.  
		Consequently, appropriate measurement cost reduction schemes\cite{Loaiza2023, Izmaylov2019, Izmaylov2019b, Yen2021, Verteletskyi2020, Yen2020} can be substantially lower the overhead due to the 3-body terms. 
		Additionally, one can ameliorate the quantum computing measurement cost problem with approaches like informationally-complete positive operator-valued measures\cite{gui2021adaptive, Fischer2024, Fischer2022, Glos2022}
		classical shadows\cite{Huang2020}
		or shadow spectroscopy.\cite{Chan2022}

	Concerning the VMC optimization of the Jastrow factor, this only considers occupied orbitals in the initial HF solution. Virtual orbitals, constructed, e.g., from commonly used correlation-consistent basis sets~\cite{Dunning_1989} are not optimized for the TC method. 
	Following Refs.~\cite{Khn2019, Verma2021, Gonthier2022} we will therefore use pre-optimized natural orbitals (NOs) from second-order M\o{}ller-Plesset (MP2) perturbation theory calculations. 
	In particular, orbital pre-optimization works exceptionally well in conjunction with the TC method by efficiently truncating the virtual orbital space and reducing the resource (qubit) requirements further. 
	(see the Methods Section~\ref{sec:methods} and the SI\cite{SI} for details).
	For a detailed comparison between the use of HF orbitals and MP2-NOs for LiH calculations see the Sections IC and  II of the SI \cite{SI}.
	The overall workflow of the TC-VarQITE algorithm is sketched in Fig.~\ref{fig:sketch-comb}e. 
	
		We want to summarize the additional computational cost of our proposed TC-VarQITE approach using VMC-optimized Jastrow factors and MP2-NOs compared to running VQE using HF orbitals. 
		The baseline cost of VQE+HF is as a formally quartic scaling of HF with the number of orbitals, $\mathcal{O}(N^4)$,\nocite{Helgaker} (although practically the cost is usually closer to $\mathcal{O}(N^3)$) and a quantum measurement scaling of VQE of $\mathcal{O}(N^4)$. 
		These estimates assume \enquote{vanilla} implementations of HF (ignoring optimized implementations,  i.e. using local approximation\nocite{Schwegler1996, Ochsenfeld1998, Kussmann2013, Goletto2021} or density-fitting\nocite{Kppl2016}) and VQE (with no gradient information,\nocite{Stokes2020} optimized classical optimizers\nocite{Fitzek2023, Tilly2022}, or advanced grouping and measurement strategies\nocite{Huggins2021, Loaiza2023, Hamamura2020, Huang2020}).
		The additional cost of our proposed method is: 
		\begin{itemize}\itemsep0pt
			\item In case of using MP2-NOs: Assuming a standard implementation,  ignoring i.e. local\nocite{Saebo1993, Schtz1999, , Werner2015, Szab2021} or density fitting/resolution of identity approximation\nocite{Feyereisen1993, Vahtras1993}, 
			MP2 formally scales as $\mathcal{O}(N^5)$.\nocite{Tang1970, Helgaker_2000}
			\item The optimization of the Jastrow factor using VMC has a square scaling, $\mathcal{O}(n_e^2)$,\nocite{Assaraf2017} in memory and a cubic time scaling with the number of electrons, $\mathcal{O}(n_e^3)$\nocite{Nightingale1998, Becca2017, Ceperley1977, Foulkes2001, Haupt2023} (ignoring i.e. optimizations based on partitioning or subsampling\nocite{Bienvenu2022}). 
			\item Ignoring any approximations to the 3-body terms\nocite{Dobrautz2022, Christlmaier2023}, the calculation (time) and the storage (memory) of the TC integrals formally scale as $\mathcal{O}(N^6)$.\nocite{Cohen2019, Christlmaier2023}
			\item Measuring the metric for (TC-)VarQITE (ignoring any approximations\nocite{Gacon2023, Gacon2021, Stokes2020, Fitzek2023}) scales as $\mathcal{O}(n_\theta^2)$ and measuring the gradient (with TC 3-body terms) scales as $\mathcal{O}(n_\theta \cdot N^6)$. 
		\end{itemize}
		
		In quantum chemistry applications the number of electrons is usually (much) smaller than the number of orbitals, $n_e < N$.
		Thus, the main increase in computational cost is the $\mathcal{O}(N^6)$ scaling due to the TC 3-body terms. 
		However, as argued above and shown in Fig.~\ref{fig:sketch-comb}f, the drastic reduction in the number of necessary orbitals (up to an order of magnitude in this work, $N \rightarrow N/10$) due to the TC method outweighs this computational overhead in the range up to 1000 qubits. 
		Accurate approximations to the 3-body terms\nocite{Cohen2019, Christlmaier2023} can reduce the scaling overhead of TC-VarQITE to $\mathcal{O}(N^5)$, extending its range of advantage compared to \enquote{vanilla} VQE well beyond 1000s of qubits.

	\section{\label{sec:results}Results and Discussion}
	
	We demonstrate the advantages of the TC approach with three applications on atomic and molecular systems: the beryllium atom, and the hydrogen and lithium hydride molecules. 
	If not specified differently, simulations are performed using HF orbitals and the unitary coupled cluster singles doubles (UCCSD) Ansatz\cite{romero2018strategies, Anand2022}, which gives reasonable indications about the performance of the TC method compared to the non-TC one. 
	To demonstrate its potential, in this study, we initially solve the TC-VarQITE algorithm in the matrix formalism (statevector simulation), implying that all gates are implemented exactly (neither qubit decoherence nor gate infidelities are considered), and sampling noise is ignored. 
	Up to 12 qubits, these noise-free results were obtained by simulation of quantum hardware. In contrast, data of larger calculations were obtained with a classical solver in the form of the TC-full configuration quantum Monte Carlo (TC-FCIQMC) method.\cite{Dobrautz2019, Guther2020, Cohen2019}. 
	
	To enable hardware calculations, we also evaluate LiH with TC-VarQITE using a hardware-efficient Ansatz (HEA)\cite{Kandala2017} and compared the results with non-TC calculations. 
	Finally, to demonstrate the current and near-term hardware applicability of the TC approach, we calculate the LiH dissociation energy both with a noisy quantum circuit simulator and with actual hardware (HW) experiments on the 7-qubit \texttt{ibm\_lagos} device.
	
	Note that since the number of qubits needed to do a practical quantum computation is approximately equal to the number of spin-orbitals, in what follows we use these two terms interchangeably. Details on the basis sets used in our calculations are provided in the Figure captions and the SI.~\cite{SI}

	\subsection{\label{sec:be}Beryllium atom}
	
\begin{figure}
		\centering
		\includegraphics[width=0.45\textwidth]{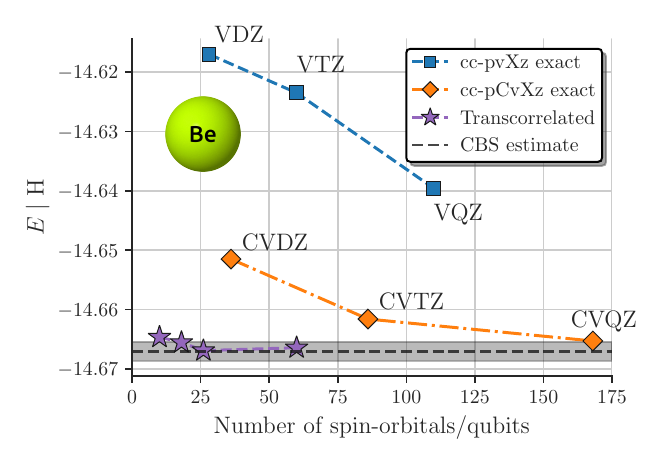}
		\vspace*{-0.5cm}
		\caption{\label{fig:results-be} 
			All-electron TC-VarQITE and non-TC FCI results for the beryllium atom using HF orbitals as a function of the number of spin-orbitals (or qubits). 
			TC-VarQITE reaches chemical accuracy (gray area) of CBS limit estimates\protect\cite{Davidson1991} (black dashed line) using only 18 qubits.}
	\end{figure}

	Figure~\ref{fig:results-be} shows all-electron TC-VarQITE results as a function of basis set size for the beryllium atom.  
	To achieve results within chemical accuracy compared to the CBS limit (i.e., 1 kcal/mol = 1.6 mHartree (the gray area in Fig.~\ref{fig:results-be}) with an FCI calculation, one would need
	168 qubits, far beyond what can currently be used efficiently. 
	The TC method, on the other hand, provides energies within chemical accuracy of the exact CBS limit while requiring only 
	18 qubits. 
	This near-CBS accuracy shows the potential of utilizing an explicitly correlated method (without any approximation) in the form of the TC approach to enable near-term quantum devices to yield accurate results for relevant quantum chemical problems. 
		We want to note that Schleich \textit{et al.}~\cite{schleich2021improving} have recently also obtained highly-accurate results for the beryllium atom using small basis sets using the approximate VQE+[2]$_{R12}$ explicitly correlated method.

	\subsection{\label{sec:h2}Hydrogen molecule}
	
\begin{figure*}
		\centering
		\includegraphics[width=\textwidth]{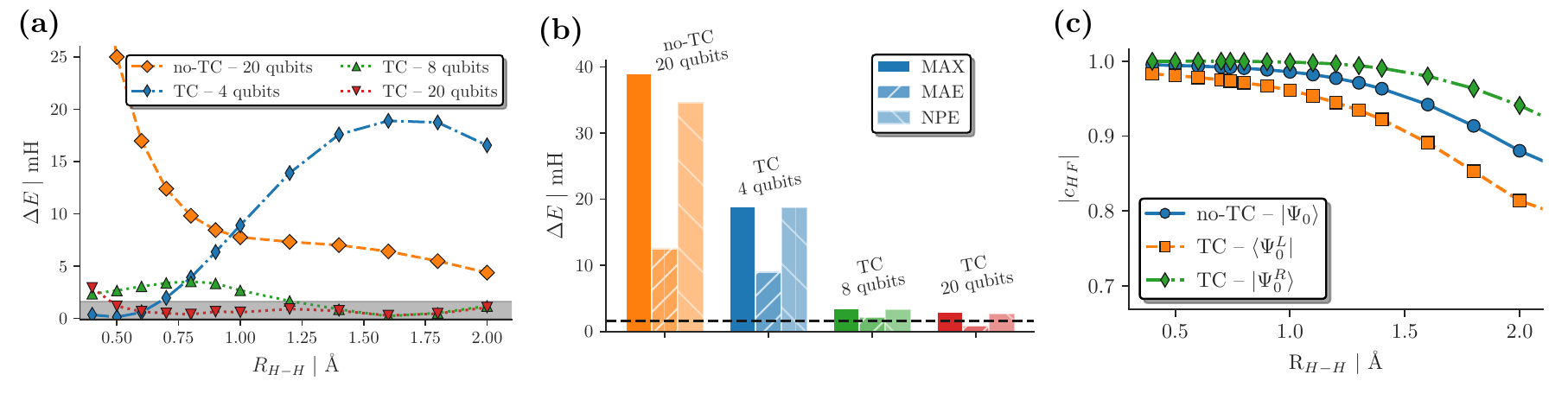}
		\vspace*{-0.7cm}
		\caption{\label{fig:results-h2}
			(a) Energy errors for TC calculations of the H$_2$ bond dissociation w.r.t. CBS limit results (mH vs. \AA). TC-VarQITE calculations are based on HF orbitals in a STO-6G (4 qubits), 6-31G (8 qubits), and cc-pVDZ (20 qubits) basis sets. Also shown are no-TC FCI/cc-pVDZ calculations.
			The grey bar indicates chemical accuracy.
			(b) Error statistics in the form of the maximum error (MAX), the mean average error (MAE), and the non-parallelity error (NPE) for calculations shown in (a).
			(c) Hartree-Fock weight in the ground state wavefunction of the original Hamiltonian (no-TC), the left, $\bra{\Psi_0^L}$ and right, $\ket{\Psi_0^R}$, eigenvectors of the TC Hamiltonian, all computed in the cc-pVDZ basis set.
		}
	\end{figure*}

	Figure~\ref{fig:results-h2}a compares TC-VarQITE results for the H$_2$ bond dissociation with CBS limit results.
	We also compare with conventional FCI calculations in a cc-pVDZ basis set (corresponding to 20 qubits). 
	TC-VarQITE results are shown for 
	increasing basis set sizes using 4, 8, and 20 qubits, respectively. 
	TC-VarQITE allows
	near chemically accurate results (with respect to the CBS limit, c.f., gray area in Fig.~\ref{fig:results-h2}a)  across the entire binding curve using only 
	8 qubits.
	It is noteworthy that whereas we reach near chemical accuracy with 20 qubits, conventional methods require at least 120 spin-orbitals 	for the same performance (see the SI\cite{SI}).

	The additional benefit of increased compactness of the right eigenvector of the TC Hamiltonian\cite{Dobrautz2019, Sokolov2022} can be appreciated in Fig.~\ref{fig:results-h2}c. 
	The TC right eigenvector retains a more significant Hartree-Fock weight ($c_{\rm HF}$) and, thus, single-reference character across the whole H$_2$ binding curve. 
	Note how the increase of the $c_{\rm HF}$ component is particularly pronounced relative to the original ground state (no-TC) wavefunction in the strongly correlated dissociation regime, which is challenging for standard post-HF methods. 
	Like the Hubbard model studied in Ref.~\cite{Sokolov2022}, this increased compactness results in shallower circuit Ans\"atze for the ground state wavefunction.
	
	\subsection{\label{sec:lih}Lithium hydride}
	
\begin{figure*}
		\centering
		\includegraphics[width=0.85\textwidth]{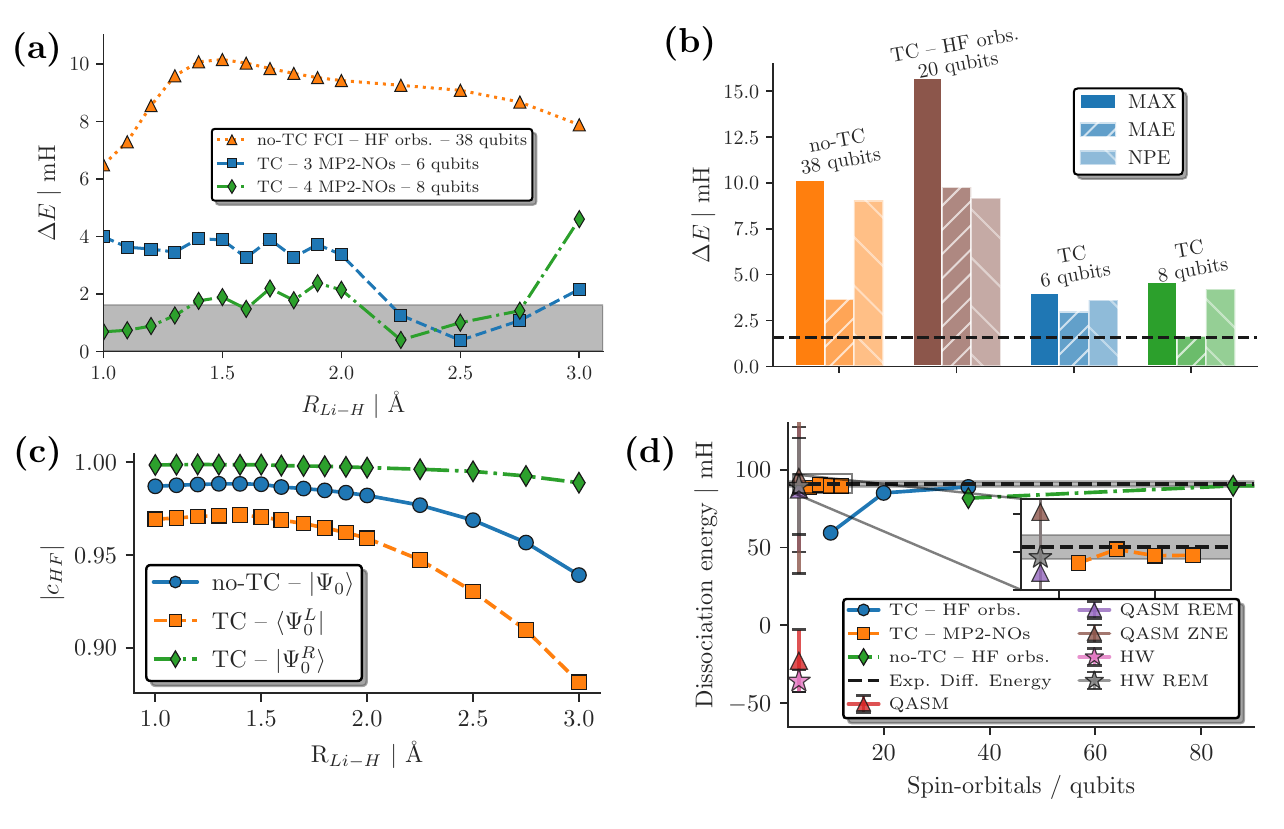}
		\vspace*{-0.6cm}
		\caption{\label{fig:results-lih}
			\textbf{(a)} Error of TC-VarQITE calculations using the three and four most occupied MP2-NOs in a cc-pVDZ basis for LiH w.r.t. CBS limit estimates in mH vs. bond distance. 
			We compare with no-TC FCI/cc-pVDZ (38 qubits) calculations  using HF orbitals.
			\textbf{(b)} MAX, MAE, and NPE values for results shown in (a).
			\textbf{(c)} Hartree-Fock coefficient, $c_{HF}$, of the all-electron ground state wavefunction using 14 MP2-NOs for the original Hamiltonian (no-TC) and the left, $\bra{\Psi_0^L}$ and right, $\ket{\Psi_0^R}$, eigenvectors of the TC Hamiltonian. Because the compactification of the right eigenvector is more pronounced for larger systems, a higher number of MP2-NOs are used to demonstrate this behavior. 
			\textbf{(d)} LiH dissociation energy estimates (in mH) obtained with the TC method using HF orbitals in an STO-6G, 6-31G, and cc-pVDZ basis set (blue circles), MP2 NOs (orange squares), and conventional no-TC calculations (green diamonds) as a function of the number of spin-orbitals/qubits compared to experiment.~\protect\cite{Stwalley1993, Rumble2022-gy}
			QASM simulations and HW experiments on the \texttt{ibm\_lagos} device are shown as triangles and stars, respectively. 
			QASM and HW calculations used  3 MP2-NOs (4 qubits with parity encoding, see circuit in Fig.~\ref{fig:lih-statevec}b. 
			Two independent error mitigation techniques (REM and ZNE) were applied to the noisy QASM/HW results. The grey bars indicates chemical accuracy. 
		}	
	\end{figure*}

	Figure~\ref{fig:results-lih}a-b show the corresponding error analysis and comparison for the LiH  molecule. 
	TC-VarQITE provides drastically improved energies  compared to conventional FCI/cc-pVDZ calculations (corresponding to 38 qubits). 
	It is striking that it manages to do that by
	using only the four most occupied MP2-NOs (see the Methods Section~\ref{sec:methods} and the SI\cite{SI}),
	requiring only 8 qubits on quantum hardware. 
	TC-VarQITE yields results within or near chemical accuracy w.r.t. CBS limit (c.f., gray area in Fig.~\ref{fig:results-lih}a) across the whole binding curve. 
	The statistical error analysis shown in Fig.~\ref{fig:results-lih}b demonstrates how with just 3 or 4 MP2 NOs (corresponding to 6 and 8 qubits using a Jordan-Wigner Fermion-to-qubit encoding, respectively) 
	TC-VarQITE readily outperforms conventional methods, even when these are leveraging more orbitals. 
	We want to note that recently Motta \textit{et al.}~\cite{Motta2020} and Kumar \textit{et al.} obtained highly-accurate results for H$_2$ and LiH using small basis sets using the approximate CT-F12 explicitly correlated method.
	
\begin{figure}
		\centering
		\includegraphics[width=0.5\textwidth]{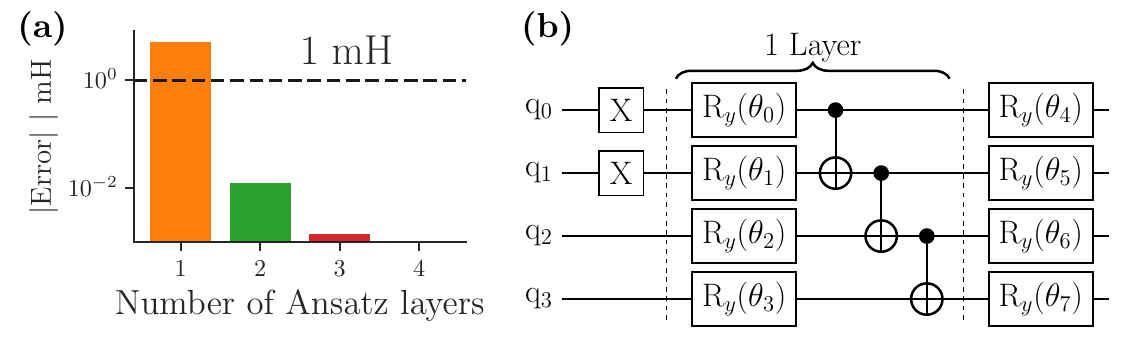}
		\caption{\label{fig:lih-statevec} 
			\small
			All-electron TC-VarQITE statevector simulations of LiH at equilibrium bond distance with 3 MP2-NOs. 
			(a) Error (in mH) of the $R_y$-Ansatz simulations w.r.t. the TC-FCI energy as a function of the number of used layers. 
			Two layers of the $R_y$-Ansatz suffice to obtain results within 1 mH of TC-FCI/cc-pVDZ using 3 MP2 NOs.
			(b) $1$-layer $R_y$-Ansatz with linear entanglement and a final rotation layer.
		}
	\end{figure}

	We note that (in Figure~\ref{fig:results-lih}c)  the resulting \enquote{compactification} of the wavefunction (and corresponding circuit) is much more pronounced for LiH than for H$_2$. 
	This increased compactness suggests an increasing benefit of the TC approach for larger systems and exemplifies the favorable scalability of the method. 
	With an HF coefficient greater than 0.99 over the entire dissociation profile, the TC right eigenvector can be efficiently mapped to exceedingly shallow quantum circuits suited for hardware calculations, as shown in Fig.~\ref{fig:lih-statevec}a.

	With a TC Hamiltonian, we can calculate the dissociation energy of LiH with \textit{experimental} accuracy with less than ten qubits (Figure~\ref{fig:results-lih}d), a hardware cost that is compatible with experiments on current and near-term quantum devices. 
	In contrast, no-TC methods would require a basis set as large as cc-pVTZ, corresponding to 88 spin-orbitals, to reach comparable results, as shown in Fig.~\ref{fig:results-lih}d.
	
	To further substantiate the near-term potential of TC-VarQITE, we study the equilibrium bond distance of LiH using 3 MP2-NOs and a hardware-efficient Ansatz.~\cite{Kandala2017} 
	In this instance, we use repeated layers of $R_y$ rotational gates (applied to each qubit) followed by linear entangling \texttt{CNOT} gates (see Fig.~\ref{fig:lih-statevec}b). 
	Parity encoding and 2-qubit reduction are also applied in this example. 
	
	Errors of this approach w.r.t. exact (state vector) UCCSD results are shown as a function of the number of Ansatz layers in Figure~\ref{fig:lih-statevec}a.
	Already with two layers (16 single qubit  R$_y$ gates and 6 CNOT gates), results are within $10^{-3}$ Ha from the UCCSD reference. 
	To bring this result into perspective, not even a full cc-pVDZ basis (36 qubits with parity reduction) calculation would enable this level of accuracy with conventional methods (see Fig.~\ref{fig:results-lih}c).

	To test the hardware (HW) applicability of TC-VarQITE, we have applied it to calculate the LiH dissociation energy, which is known experimentally (Table~\ref{tab:spect-const}). 
	This calculation was done using a 1-layer version of the hardware-efficient Ansatz shown in Fig.~\ref{fig:lih-statevec}b, while initializing in the HF state, $\ket{\Phi_{\rm HF}}$, first in QASM simulations, then on HW (further details are provided in the SI). 
	To account for the effect of noise, which causes raw QASM/HW results to be widely off the mark (Fig.~\ref{fig:results-lih}d), error mitigation was applied.  
	We have separately tested two techniques: zero noise extrapolation (ZNE)\cite{Temme2017, Li2017} and reference error mitigation (REM),~\cite{Lolur2023} both alongside readout error mitigation, details of which can be found in the SI.~\cite{SI}
	Even though the standard deviations of our HW results are sizeable due to noise, with error mitigation, TC-VarQITE yields QASM and HW predictions of the LiH dissociation energy close to (HW+ZNE) or within (HW+REM) chemical accuracy of the experimental results. 

	\begin{table}
		\centering
		\small
		\caption{\label{tab:spect-const} Comparison with experimental data.~\protect\cite{Huber1979, Stwalley1993} Equilibrium distances ($R_e$), dissociation energies ($D_0$), and vibrational frequencies ($\omega_e$) are shown for H$_2$ and LiH with and without TC. 
			Results closest to $R_e$ experimental data by Motta \textit{et al.}~\cite{Motta2020} using the CT-F12 method are also reported.}
	\begin{tabular}{ccccc} 
		\toprule
		& & \multicolumn{3}{c}{H$_2$} \\  
		\cline{2-5}
		&	Qubits$^a$ & $R_e$(\AA) & $D_0$(eV) & $\omega_e$(cm$^{-1}$) \\ 
		\hline
		\multirow{3}{*}{no-TC$^b$} & 2 & 		0.7330 & 3.67 & 4954  \\ 
		& 6 & 		0.7462 & 3.87 & 4297 \\ 
		& 18 & 		0.7609 & 4.19 & 4353 \\ 
		\hline
		CT-F12$^e$ & 6 & 0.7397 &  -- & 4462 \\
		\hline
		\multirow{2}{*}{TC$^c$} &  \bf 2 & \bf 0.7346 & \bf 4.69 & \bf 4435 \\ 
		&  \bf 6 &  \bf 0.7428 &  \bf 4.66 &   \bf 4361  \\
		\hline
		Exp. &  & \bf 	0.7414 & \bf 4.52& \bf 4401 \\
		\midrule[1pt]
		&  & \multicolumn{3}{c}{LiH} \\
		\cline{2-5}
		&	qubits & $R_e$(\AA) & $D_0$(eV) & $\omega_e$(cm$^{-1}$) \\ 
		\hline
		\multirow{3}{*}{no-TC$^b$} 	& 10 & 1.5422 &  2.66 & 1690 \\
		&  20 & 1.6717 &  1.80 & 1283 \\
		& 36 & 1.6154 &   2.17& 1360 \\
		\hline
		CT-F12$^f$ & 18 & 1.615\phantom{0} &  -- & 1385 \\
		\hline
		\multirow{2}{*}{TC$^d$} & \bf 4 & \bf 1.6032  & \bf 2.42 & \bf 1377 \\
		&   	 \bf 6 &  \bf 1.5998 &  \bf 2.47  &   \bf 1390 \\
		\hline
		Exp. & & \bf  1.5949 & \bf  2.47 &\bf  1405  \\
		\bottomrule
	\end{tabular}\\
	{\footnotesize $^a$With parity encoding and 2-qubit reduction. $^b$STO-6G, 6-31G, and cc-pVDZ basis sets. $^c$STO-6G and 631-G basis. $^d$3 and 4 MP2-NOs. For more details, see the SI.~\cite{SI} $^e$q-UCCSD/6-31G results of Ref.~\cite{Motta2020}. $^f$q-UCCSD/comp results of Ref.~\cite{Motta2020}.} 
\end{table}

\subsection{Comparison with experimental data and quantum hardware requirements}
To further evaluate the TC-VarQITE approach, we calculated equilibrium bond lengths, $R_e$, and vibrational stretching frequencies, $\omega_e$, (in addition to the above-studied dissociation energies, $D_0$) for the H$_2$ and LiH molecules 
and benchmark it against available experimental data (Table~\ref{tab:spect-const}) 
as well as highly-accurate CT-F12 results by Motta \textit{et al.}~\cite{Motta2020}
We note excellent agreement for all spectroscopic quantities obtained with TC-VarQITE using only two qubits for H$_2$ 
and four qubits for LiH 
and consistently equal good results using 6 qubits
(see the SI\cite{SI} for details).

Estimates on the necessary quantum hardware requirements to obtain the results of Table~\ref{tab:spect-const} with TC-VarQITE are summarised in Table~\ref{tab:depth}. 
We report the number of Ansatz parameters, the number of CNOTs, the total number of (1- and 2-qubit) gates, and the circuit depth -- the number of quantum gates that cannot be executed simultaneously. 
We also show selected estimates of calculations without transcorrelation (no-TC) and available data by Motta \textit{et al.}\cite{Motta2020} using the CT-F12 approach. 
All our estimates use parity Fermion-to-qubit encodings (with a subsequent 2-qubit reduction) and all but one use the default UCCSD implementation of Qiskit\cite{qiskit}. 
The entry indicated by TC+HEA employs a 2-layer hardware efficient R$_y$ Ansatz with linear entangling shown in Fig.~\ref{fig:lih-statevec}b that yields sub microhartree precision for LiH at equilibrium bond distance (Fig.~\ref{fig:lih-statevec}a)

The results of Table~\ref{tab:depth} demonstrate the drastic reduction in necessary quantum resources, not only in the total number of qubits but also in the required circuit depth. 
TC-VarQITE results for H$_2$ using an STO-6G basis (two qubits with parity encoding) need only 4 CNOT gates and a circuit depth of 14 to yield results closer to experiment than no-TC or CT-F12 calculations\cite{Motta2020} requiring over 400 CNOTs. 
The powerful combination of TC with MP2-NOs is demonstrated for LiH. 
TC-VarQITE using 3 MP2-NOs (4 qubits with parity encoding) requires only 172 CNOTs and a circuit depth of 275 to yield highly accurate spectroscopic data. 
Alternative approaches (no-TC or CT-F12) need larger basis sets, more qubits, and much deeper circuits to achieve similar accuracy. 
Using hardware-efficient Anätze further drastically reduces the TC-VarQITE hardware requirements to only 6 CNOT gates and a quantum circuit depth of 10. 

On the other hand, Table~\ref{tab:depth} also demonstrates the drawback of the TC approach in the form of the increased number of Pauli terms due to the 3-body in the Hamiltonian, Eq.~\eqref{eq:tc}.
I.e. the H$_2$ TC Hamiltonian using a 6-31G basis (6 qubits with parity encoding) has 607 terms compared to the 235 terms of the CT-F12 and 159 terms of the original (no-TC) Hamiltonian. 
However, with TC-VarQITE, using an STO-6G basis, and thus only 7 Pauli terms, suffices to reach the same accuracy as other methods in larger basis sets.

\begin{table}
	\centering\small 
	\caption{\label{tab:depth}
			Estimate of quantum circuit requirements for the calculation of spectroscopic constants in Table~\ref{tab:spect-const} using parity encoding with 2-qubit reduction and a default UCCSD Ansatz (except the last row). 
			Number of parameters of the quantum circuit Ansatz, number of two-qubit CNOTs and the total number of gates (obtained with Qiskit's \texttt{count\_ops()} function), as well as circuit depth (sets of quantum gates that cannot be executed simultaneously).
			Data for CT-F12 calculations are taken from Ref.~\cite{Motta2020}
	}
	\begin{tabular}{cccccccccc}
		\toprule
		System & Basis &    Orbitals & Qubits & Method & Paulis$^a$ & Parameters & Gates & CNOTs & Depth \\
		\midrule
		\multirow{5}{*}{\normalsize H$_2$}  & \multirow{2}{*}{STO-6G} &    \multirow{2}{*}{2}	& \multirow{2}{*}{2} &  no-TC	&	5	& \multirow{2}{*}{3}  	& \multirow{2}{*}{21}  	& \multirow{2}{*}{4} & \multirow{2}{*}{14}   \\ 
		&\bf  	& 	& 	& \bf TC &  \bf 	7	&   	&   	& &    \\ 
		\cline{2-10}
		& \multirow{3}{*}{6-31G} &  \multirow{3}{*}{4}	& \multirow{3}{*}{6}	&  	no-TC & 159	&  \multirow{3}{*}{15}  	& 1271  & 560 & 779   \\ 
		&	  & & 	& 	\bf TC &  \bf 	607	& 	&  1271  		&   560 &  779   \\ 
		&    &  &  	& CT-F12$^b$	& 235 		&  	& 741 		& 476 & 604  \\ 
		\hline
		\multirow{4}{*}{\normalsize LiH} &  cc-pVDZ & 18 & 36 &  no-TC$^b$  & -- & 323  & 110230 &  89080 & 95507 \\ 
		&  6-31G & 10 	& 18 & CT-F12$^b$  	& 8527 	& 99 	& 12087 & 9644 & 10780 \\
		\cline{2-10}
		& \multirow{2}{*}{\bf   MP2-NOs} & \multirow{2}{*}{\bf  3} & \multirow{2}{*}{\bf 4} &\bf TC   & \multirow{2}{*}{\bf  108}	& \bf 8 &\bf 430 &\bf  172 &\bf  275  \\			
		&  	 & &  & \bf TC+HEA$^c$ &  &\bf  12 & \bf 20 & \bf 6 &\bf  10 \\
		\bottomrule
	\end{tabular}\\
	{\footnotesize 	
			$^a$Terms smaller than $10^{-8}$ Ha in absolute value are omitted. 
			$^b$From Ref.~\cite{Motta2020}. 
			$^c$Using a 2-layer hardware efficient R$_y$ Ansatz with linear entangling shown in Fig.~\ref{fig:lih-statevec}b that yields sub microhartree precision for LiH at equlibrium bond distance (Fig.~\ref{fig:lih-statevec}a)}
\end{table}

\section{\label{sec:discussion} Conclusions and Outlook}

This article describes a quantum computing implementation of an explicitly correlated method based on the transcorrelated (TC) approach.
The TC method drastically reduces the number of required qubits and circuit depth to obtain results within chemical accuracy to experiment. 
Here we consider the exact TC formalism and propose efficient theoretical and computational solutions to overcome the challenges 
preventing 
its implementation 
on near-term quantum computers. 

By incorporating the electron cusp condition, the TC method approaches complete basis set limit results and enables chemically accurate calculations with relatively small basis set sizes.
Because the TC Hamiltonian is non-Hermitian, 
it cannot be directly combined with the conventional variational quantum eigensolver. 
To overcome this issue, we made use of the variational (Ansatz-based) quantum imaginary time evolution algorithm (VarQITE),~\cite{mcardle2019variational} for which recent advances~\cite{Sokolov2022} enable an efficient extension to non-Hermitian problems. 
In addition, we employ a pre-optimized set of natural orbitals obtained from second-order M\o ller-Plesset perturbation theory calculations~\cite{Khn2019} (MP2-NOs) to efficiently truncate the virtual orbital space. 
MP2-NOs work exceptionally well in conjunction with the TC method and help to further reduce the number of required qubits. 

We demonstrate the TC-VarQITE approach, combined with orbital optimization, 
on small atomic and molecular test systems, including the beryllium atom, the hydrogen dimer, and lithium hydride. 
In all these cases, we could closely reproduce experimental values, including bond lengths, dissociation energies, and the vibrational frequencies of H$_2$ and LiH, using just two and four qubits, respectively.
Finally, to illustrate the applicability of the TC-VarQITE approach in quantum hardware experiments, we also evaluated the bond dissociation energy of LiH. 
When combined with error-mitigation techniques, our hardware results show a great level of accuracy, close to the CBS limit and spectroscopic data.
The mitigation techniques include 
zero noise extrapolation,~\cite{Temme2017, Li2017} reference-state error mitigation,~\cite{Lolur2023} together with the commonly used readout error mitigation.~\cite{bravyi2021mitigating}

The aim of this work was the implementation and demonstration of the prowess of  un-approximated TC approach to \textit{ab initio} molecular on quantum hardware. 
To do this, we chose what might be considered \enquote*{minimal} test systems.
However, as has been done on \enquote{conventional} hardware,~\cite{Cohen2019, Dobrautz2022, Haupt2023, Christlmaier2023, Schraivogel2023, Ammar2022} in future work, we will extend the application of TC approach to larger molecular systems than studied here. 
Additionally, we will develop new methodologies to obtain not only energy estimates, but also properties in form of unbiased density matrices, and consequently, combine the TC approach with self-consistent orbital optimization~\cite{Roos1980, Olsen2011, Dobrautz2021, fitzpatrick2022selfconsistent, deGraciaTrivio2023}/embedding,~\cite{Bauer2016, Tilly2021, Rossmannek2023, Rossmannek2021} spin-conserving schemes,~\cite{Dobrautz2019b, Dobrautz2022b, Yun2021, LiManni2021, LiManni2020} 
as well as adaptive quantum circuit Ansätze.~\cite{Grimsley2019, tang2021qubit, Gomes2021, Magnusson2024}

In conclusion, the full potential of the TC method manifests as a dramatic cost saving (in terms of the number of qubits and circuit depth) for current quantum hardware calculations. 
Our study demonstrates that the TC-VarQITE has the potential to become the method of choice for calculating accurate quantum chemistry observables of relevant molecular systems on current and near-term quantum computers.

\section{Author Contributions}
Conceptualization: WD, IS, AA and IT.
Data curation: WD.
Formal Analysis: WD.
Funding acquisition: WD, MR, AA and IT.
Investigation: WD.
Methodology: WD, PR, AA and IT.
Project Administration: WD, MR, AA and IT.
Resources: MR, IT and AA.
Software: WD, KL, IS, and PR.
Supervision: MR, AA and IT.
Visualization: WD.
Validation: WD.
Writing -- original draft: WD, MR, IT and AA.
Writing -- review and editing: WD, IS, KL, PR, MR, AA and IT.

\section{Conflicts of interest}

There are no conflicts to declare.

\section{Acknowledgments}

Funded by the European Union. 
Views and opinions expressed are, however, those of the author(s) only and do not necessarily reflect those of the European Union or REA. 
Neither the European Union nor the granting authority can be held responsible for them.
This work was funded by the EU Flagship on Quantum Technology HORIZON-CL4-2022-QUANTUM-01-SGA project 101113946 OpenSuperQPlus100. 
This research has been supported by funding from the Wallenberg Center for Quantum Technology (WACQT).
WD acknowledges funding from the European Union’s Horizon Europe research and innovation programme under the Marie Sk{\l}odowska-Curie grant agreement No. 101062864. 
IT acknowledges funding from the NCCR MARVEL, a National Centre of Competence in Research by the Swiss National Science Foundation (grant number 205602). 
This research relied on computational resources provided by the National Academic Infrastructure for Supercomputing in Sweden (NAISS) at C3SE and NSC, partially funded by the Swedish research council through grant agreement no 2022-06725.
We acknowledge the use of IBM Quantum services for this work. 
The views expressed are those of the authors and do not reflect the official policy or position of IBM or the IBM Quantum team.
IBM, the IBM logo, and ibm.com are trademarks of International Business Machines Corp., registered in many jurisdictions worldwide.
Other product and service names might be trademarks of IBM or other companies. 
The current list of IBM trademarks is available at \href{https://www.ibm.com/legal/copytrade.}{https://www.ibm.com/legal/copytrade.}

The authors declare no competing interests.

\section{Additional Information}

Supporting Information  (SI) is available for this paper.\cite{SI}
The SI contains further details and results concerning the computational workflow and settings for the beryllium, H$_2$ and LiH calculations; the use of MP2 natural orbitals; the H$_2$ CBS limit extrapolation; the calculation of the LiH dissociation energy and spectroscopic constants; the QASM simulations / quantum hardware calculations and error mitigation schemes; the 1-norm  of the transcorrelated qubit Hamiltonian. 
This information is available free of charge via the Internet at \url{http://pubs.acs.org}

Data and software to reproduce this work  is available in an accompanying public Github repository.\cite{github-repo}

\section{\label{sec:methods}Methods}

\subsection{\label{sec:tc}Transcorrelation}

Transcorrelation is the application of a similarity transformation to the Schr\"odinger equation of a system, $\hat H \Psi = E \Psi$, to absorb the Jastrow factor $e^{\hat J}$ from the Ansatz $\Psi = e^{\hat J}
\Phi$ into an effective Hamiltonian. 
The resulting TC Schr\"odinger equation, $\hat{H}_{\rm TC} \Phi = E \Phi$, can be solved in second quantization using any quantum chemistry eigensolver, including quantum computers, with the advantage
that the FCI solution for $\Phi$ is much more compact than that for $\Psi$, and thus easier to represent approximately.
Eigensolvers only require the values of the matrix elements of ${\hat H}_{\rm TC}$ between different determinants.
If the Jastrow factor can be written as $J = \sum_{i<j} u({\bf r}_i, {\bf r}_j)$ then
\begin{equation}
\hat{H}_{\rm TC} =
\hat{H}
- \sum_{i<j}\hat{K}({\bf r}_{i},{\bf r}_{j})
- \sum_{i<j<k} \hat{L}({\bf r}_{i},{\bf r}_{j},{\bf r}_{k}) \;,
\end{equation}
where $\hat K$ is an operator that modifies the values of two-electron matrix elements and introduces non-Hermiticity and $\hat L$ is an operator that connects determinants separated by triple excitations.
Eigensolvers thus need the ability to accommodate non-Hermiticity and three-electron matrix elements, so non-TC implementations usually require some degree of modification.

We use a Drummond-Towler-Needs Jastrow factor,~\cite{Drummond2004, Lopez2012}
which we optimize with variational Monte Carlo (VMC)\cite{Ceperley1977, Foulkes2001, Haupt2023} (with a scaling of $\mathcal{O}(n_e^3)$ on conventional hardware) using the \texttt{CASINO} package.\cite{Needs2020, LopezRios2006}
We optimize the Jastrow factor by minimizing the variance of the TC reference energy, as proposed recently in Ref.~\citen{Haupt2023}.
We then use the \texttt{TCHint} library to calculate the 2- and 3-body integrals required to construct the TC molecular Hamiltonian in second quantization. 
See Ref.~\cite{Haupt2023} and the SI for more details and sample input files of the VMC optimization and integral calculation can be found in the \texttt{Github} repository accompanying the paper.\cite{github-repo}

\subsection{\label{sec:qite}Variational Ansatz-based quantum imaginary time evolution}

The VarQITE algorithm~\cite{mcardle2019variational} is based on McLachlan's variational principle, which is used to derive the evolution of gate parameters, represented by $\boldsymbol{\theta}(\tau)$, for a wavefunction Ansatz.
The derivation is encapsulated in Eq.~(4) of the main text, which leads to a linear system of equations defined in Eq.~5 of the main text. 
This system necessitates the computation of matrix elements associated with the matrix $\mbf A$ and the gradient vector $\mbf C$ defined as 
\begin{equation}
\label{eq:A}
\begin{aligned}
	A_{ij} &= \Re\left(\frac{\partial \bra{\Phi(\boldsymbol{\theta}(\tau))}}{\partial \theta_i}\frac{\partial \ket{\Phi(\boldsymbol{\theta}(\tau))}}{\partial \theta_j}\right) \, ,
\end{aligned}
\end{equation}
and
\begin{equation}
\label{eq:C}
\begin{aligned}
	C_i =& \Re\left(\frac{\partial \bra{\Phi(\boldsymbol{\theta}(\tau))}}{\partial \theta_i} \hat{H}\ket{\Phi(\boldsymbol{\theta}(\tau))}\right), \\
\end{aligned}
\end{equation}
where the wavefunction Ansatz is differentiated with respect to the gate parameters. 
In our implementation, their calculation is performed using the methodology outlined in,~\cite{Sokolov2022} specifically designed for non-Hermitian (TC) problems.
Next, we give more details about the steps necessary to reproduce the results of this work.

In numerical simulations, the values of $A_{ij}$ and $C_i$ are estimated using the forward finite-differences method~\cite{milne2000calculus} given by
\begin{equation}
\frac{\partial \ket{\Phi(\boldsymbol{\theta})}}{\partial \theta_j} \approx \frac{\ket{\Phi(\boldsymbol\theta + \Delta \bold{\hat{e}}_j)} - \ket{\Phi(\boldsymbol\theta)}}{\Delta},  
\end{equation}
where $\bold{\hat{e}}_j$ is $j$-th element of the $n_\theta$-dimensional unit vector. We chose a step-size of  $\Delta = 10^{-3}$ in this work. 
To generate the state-vector representation of the Ansatz, $\ket{\Phi(\boldsymbol{\theta})}$, we create the corresponding quantum circuit in Qiskit~\cite{qiskit} and then convert it to a state-vector. 
This approach allows for the incorporation of gate errors through realistic noise models of IBM Quantum processors.
The matrix elements $A_{ij}$ and $C_i$ can be computed independently, and we parallelize their computation on multiple CPUs using the \texttt{ipyparallel} library to speed up our simulations. 
Although the forward finite-differences method provides satisfactory results for the computation of the derivatives, the parameter-shift rule~\cite{schuld2019evaluating} can also be employed within our framework to obtain analytic derivatives.

In hardware calculations, the matrix elements $A_{ij}$ and $C_i$ are calculated via the differentiation of general gates by means of a linear combination of unitaries.~\cite{schuld2019evaluating}
To compute the $C_i$ elements, we use the quantum circuit shown in Fig.~1.
For a Hermitian Hamiltonian, a Hadamard gate (H) is applied before measuring the ancilla.
For a non-Hermitian Hamiltonian, $\bar H$, we decompose $\bar H$ into hermitian and anti-hermitian components denoted by $\bar{H}= (\hat{H}^{+} + \hat{H}^{-}) / 2$, where $\hat{H}^{+} = \bar{H} + \bar{H}^{\dagger}$ and $\hat{H}^{-}= \bar{H} - \bar{H}^{\dagger}$.
Subsequently, the circuit from Fig.~1 is applied to each term of $\hat{H}^{+}$ and $\hat{H}^{-}$, 
where an R$_x(\frac{\pi}{2})$ rotational gate is applied instead of a Hadamard in the case of $\hat H^-$. 
This circuit's measurement outcomes are combined to obtain $C_i$ as in Refs.~\cite{schuld2019evaluating, Sokolov2022}
To compute the $A_{ij}$ matrix elements, we proceed in the standard way, which can be found in Refs.~\cite{mcardle2020quantum} since they are independent of the Hamiltonian.
We typically use $10^4$ to $3.2\cdot 10^4$ shots (measurements) to collect enough statistics to accurately estimate the expected values.

For the representations of Ansatz circuits, we use Qiskit's implementation of UCCSD and hardware-efficient Ansätze with the default settings.

Having all the necessary quantities, the linear system in Eq.~5 of the main text can be approximately inverted to obtain $\dot{\boldsymbol{\theta}} = - \mbf{A}^{-1} \mbf C \,$ using the least-square solver (default settings) implemented in SciPy.~\cite{scipy}
Finally, the updated parameters are obtained from $\boldsymbol{\theta}(\tau + \Delta\tau) = \boldsymbol{\theta}(\tau) + \Delta\tau \dot{\boldsymbol{\theta}}$ for a chosen time-step of $\Delta\tau = 0.05$ in this work.

\begin{figure}[h!]
\centering
\begin{minipage}[c]{0.45\textwidth}
	\includegraphics[width=0.9\linewidth]{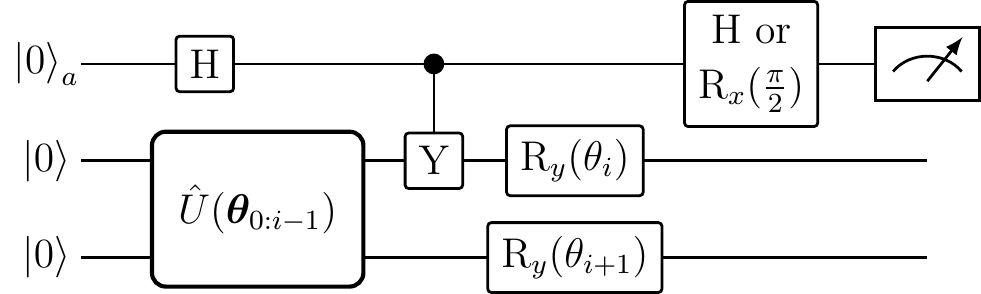}
\end{minipage}
\begin{minipage}[c]{0.45\textwidth}
	\caption{Quantum circuit used to calculate the $C_i$ term in the (non-)Hermitian case for a (TC) Hamiltonian. A Hamiltonian is first separated into its Hermitian and anti-Hermitian parts. 
		The circuit uses a Hadamard gate, H, for the Hermitian part and an R$_x(\frac{\pi}{2})$ rotational gate for the anti-Hermitian part before measuring the ancilla. This circuit needs to be repeated for every term in a Hamiltonian.}
	\label{fig:circuit_grad}
\end{minipage}
\end{figure}

Fig.~\ref{fig:circuit_grad} shows the quantum circuit used to calculate the $C_i$ term in the (non-)Hermitian case for a (TC) Hamiltonian.

\subsection{\label{sec:mp2-nos}Second Order M{\o}ller-Plesset Natural Orbitals (MP2-NOs)}

The 1-body reduced density matrix (1-RDM) in a second-quantized basis is defined as
\begin{equation}
D^p_q = \braket{\Psi|\hat{a}^{\dagger}_p\hat{a}_q|\Psi},
\label{equ:1rdm}
\end{equation}
where $\ket{\Psi}$ is the wavefunction and $\hat a^{(\dagger)}_{q(p)}$ 
is the fermionic annihilation(creation) operator of an electron in orbital $q(p)$ of the current basis. The diagonalization of Eq.~(\ref{equ:1rdm}) provides eigenvalues in terms of the occupation numbers and eigenvectors that correspond to the transformation matrix from the current basis to the natural orbital basis.
L\"owdin~\cite{Lowdin1955} first used NOs to accelerate the convergence of configuration interaction calculations by retaining only those NOs with significantly nonzero occupation numbers. 
The specific natural orbitals used in this work are obtained on the MP2 level. 
First, a mean-field Hartree-Fock (HF) solution to the system under study is obtained in a reasonably large basis set, e.g., cc-pVDZ or cc-pVTZ. In the HF canonical orbital basis, the MP2 wavefunction is 
\begin{equation}
\ket{\Psi_{\rm MP2}}= \ket{\Psi_{\rm HF}} 
+ \sum_{i>j,a>b}\frac{\braket{ab||ij}}{\Delta^{ab}_{ij}}\ket{\Psi^{ab}_{ij}},
\end{equation}
where we follow the convention to use $a,b,\dots$ and $i,j,\dots$ to indicate the unoccupied (virtual) and occupied spin-orbitals, respectively. 
The antisymmetrized Coulomb integrals are defined as $\braket{ab||ij}=\braket{ab|ij}-\braket{ab|ji}$ and the denominator is $\Delta^{ab}_{ij}=\varepsilon_i+\varepsilon_j-\varepsilon_a-\varepsilon_b$ with $\varepsilon$ denoting orbital energies (diagonal elements of the Fock matrix). 

Plugging $\ket{\Psi_{\rm MP2}}$ into Eq.~(\ref{equ:1rdm}), we find 
\begin{equation}
\begin{aligned}
	D^i_j &= \delta_{i,j}  + \frac{1}{2}\sum_{kab}\frac{\braket{ki||ab}\braket{ab||kj}}{\Delta^{ab}_{ki}\Delta^{ab}_{kj}},  \qquad
	D^a_b & = \frac{1}{2}\sum_{ijc}\frac{\braket{ij||ac}\braket{bc||ij}}{\Delta^{ac}_{ij}\Delta^{bc}_{ij}},
\end{aligned}
\end{equation}
where we ignore orbital rotations between the occupied and virtual space by setting the occupied-virtual block $D^a_i = D^i_a = 0$. 
In literature,\cite{Khn2019, Verma2021, Gonthier2022} the so-called frozen natural orbitals (FNOs) are obtained by only diagonalizing the virtual-virtual block of the 1-RDM $D^a_b$. 
In this work, we diagonalize both the occupied-occupied and virtual-virtual blocks.

\end{document}